\documentclass[runningheads]{llncs}
\usepackage{graphicx}
\usepackage{amsmath}
\usepackage{algpseudocode}
\usepackage{algorithm}
\usepackage{hyperref}
\begin{document}

\title{PriorNet: lesion segmentation in PET-CT including prior tumor appearance information}
\titlerunning{PriorNet}
%
\author{Simone Bendazzoli\inst{1}\inst{2} \and
Mehdi Astaraki\inst{1}\inst{3}}
\authorrunning{S. Bendazzoli, M. Astaraki}
%
\institute{Department of Biomedical Engineering and Health Systems
, KTH Royal Institute of Technology, Stockholm , Sweden \and Department of Clinical Science, Intervention and Technology, Karolinska Institutet, Stockholm, Sweden \and Department of Oncology Pathology, Karolinska Institutet, Stockholm, Sweden }
\maketitle              
\begin{abstract}
Tumor segmentation in PET-CT images is challenging due to the dual nature of the acquired information: low metabolic information in CT and low spatial resolution in PET. U-Net architecture is the most common and widely recognized approach when developing a fully automatic image segmentation method in the medical field. We proposed a two-step approach, aiming to refine and improve the segmentation performances of tumoral lesions in PET-CT. The first step generates a prior tumor appearance map from the PET-CT volumes, regarded as prior tumor information. The second step, consisting of a standard U-Net, receives the prior tumor appearance map and PET-CT images to generate the lesion mask. We evaluated the method on the 1014 cases available for the AutoPET 2022 challenge, and the results showed an average Dice score of 0.701 on the positive cases.

\end{abstract}
\section{Introduction}
Tumor lesion segmentation is one of the primary tasks performed on PET-CT scans in oncological practice. The main aim is to identify and delineate the tumor region, enabling a quantitative assessment, performing feature extraction and planning the treatment strategy accordingly.

U-Net model \cite{ronneberger2015unet} training is the most common supervised deep learning approach that yielded promosing results for different medical image segmentation tasks \cite{brain_segmentation} \cite{ZHAO2020100357}.
We propose a 2-step approach, where the first step aims at generating and providing prior tumor appearance information to the second step, a conventional U-Net employed for the tumor segmentation.

\section{Method description}
The proposed method consists of two main modules. First, inspired by our recent Normal Appearance Autoencoder (NAA) model \cite{astaraki2022prior}, the appearance of healthy anatomies from PET-CT images is learned by training an inpainting model. In specific, a Partial Convolution Neural Network (PCNN) \cite{liu2018image} was employed to capture the distributions of healthy anatomies. Prior information regarding the appearance of the tumors was, then, estimated by calculating the residuals between the reconstructed fake healthy images from the tumoral ones. Second, the prior information highlighting the presence of candidate tumoral regions was added as an additional channel into a supervised segmentation network in order to guide the attention of the model to the candidate regions. 
\subsection{Estimating the tumor appearance}\label{tumor_appearance}
\begin{figure}[!ht]
\centering
\includegraphics[width=0.8\textwidth]{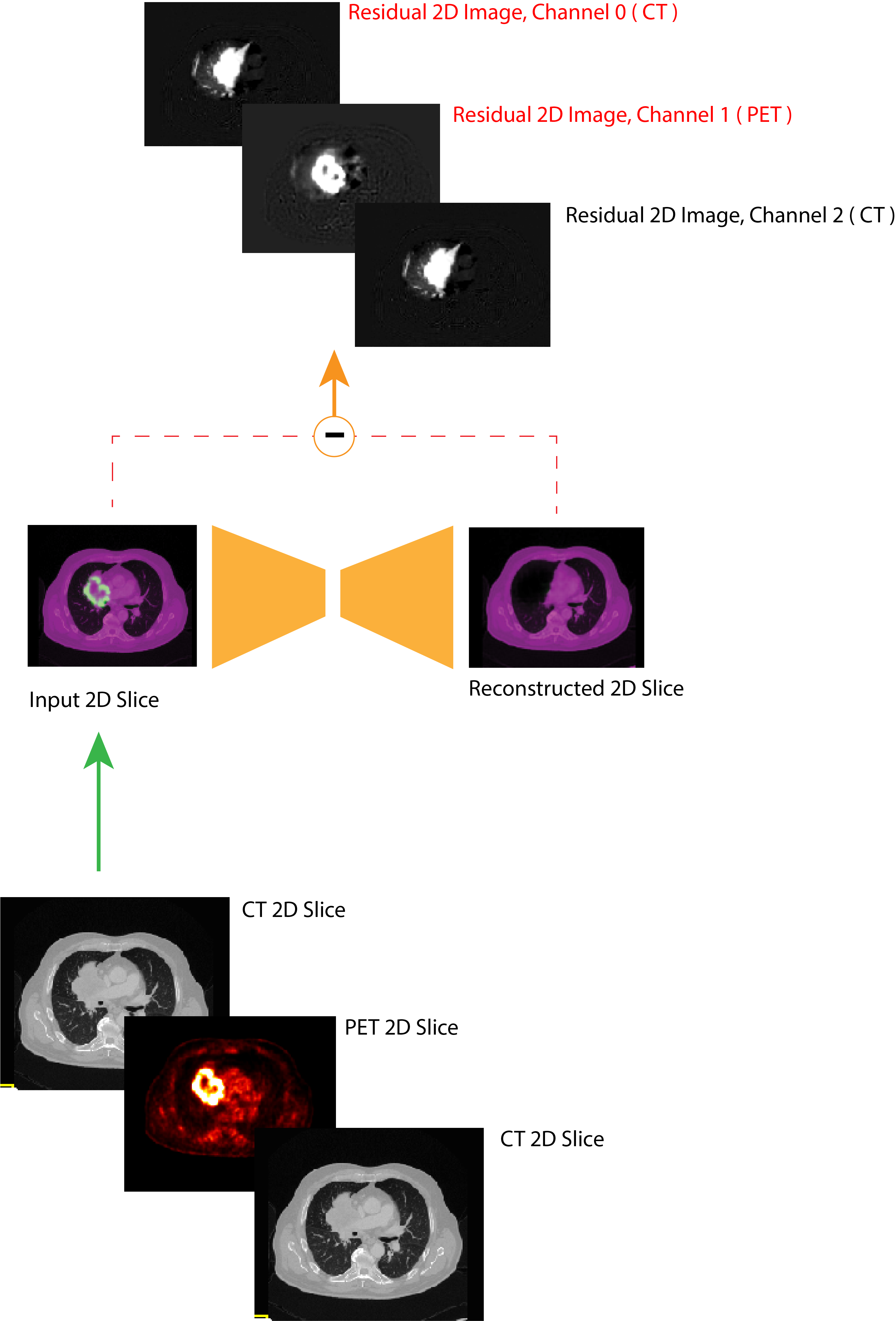}
\caption{Tumor appearance estimation workflow. An RGB 2D image is initially generated by stacking the CT and the PET slices [CT, PET, CT]. The RGB image is then reconstructed with the inpainting model. Finally, the image difference is estimated, and the first and second channels are taken as CT residual slice and PET residual slice, respectively.}
\label{fig:inpainting}
\end{figure}
Estimating the prior information regarding the appearance of tumors can be achieved by, first, modeling the healthy anatomies and then detecting the tumors as anomalies. To model the distribution of complicated healthy anatomies from whole-body PET-CT volumes, we employed a PCNN model as a robust inpainting network. This inpainting model can replace the pathological regions with the characteristics of nearby healthy tissues and generate plausible pathology-free images with realistic-looking and anatomically meaningful visual patterns. This can be achieved by the following two steps: 1) forcing the model to learn the appearance of healthy anatomies, and 2) guiding the model to inpaint only the tumoral regions. \\
To learn the attributes of healthy anatomies, healthy image slices from the PET-CT dataset were employed as the training set of the inpainting model. In specific, more than 30000 healthy image slices were used for the training set, while the pathological slices were employed for the testing set. Considering the large diversity in the shape, size, and location of the tumoral regions, random irregular shapes were synthesized by combining regular geometrical shapes, including circles, squares, and ellipses, to corrupt the healthy images. The PCNN model is trained until it fills the random holes and replaces the gaps with meaningful anatomical and imagery patterns. The objective function of the PCNN model is constructed based on several loss terms, including per-pixel loss, perceptual loss, style loss, and total variation loss. This multi-objective optimization leverage the quality of the inpainted images and reconstructs high-quality image while preserving the anatomical details. The performance of the PCNN model was evaluated by quantifying the following metrics: peak signal-to-noise ratio, mean square error, and structural similarity index. \\
During the training step, the model learns to fill the random holes with the attributes of the nearby healthy tissues. This process enforces the model to model the distribution of the healthy anatomies. Therefore, in the test phase, the learned model can be used to replace the tumoral regions with visual characteristics of healthy tissues. However, such a tumoral removal step essentially requires tumoral masks. While in the NAA model, a second autoencoder model was employed to remove the tumors automatically, in this study, we utilized the learned PCNN model to directly inpaint the tumoral regions by gaining from the hyperintensity patterns of PET images. In specific, tumoral regions in PET volumes often appear with higher FDG uptake with respect to nearby healthy tissues. Therefore, by setting a simple threshold on the PET volumes, the candidate regions for the inpainting step can be estimated. The corresponding candidate regions over the PET-CT slices are then set as a hole to be inpainted the learned PCNN model. As a result, the tumoral regions are inpainted and replaced with the appearance of learned healthy tissues.\\
It should be noted that other normal tissues with a high level of SUV can be presented in the thresholded images, and their inpainting leads to a large value of false positive rate. However, the PCNN model already learned the appearance of healthy anatomies from the training phase. Therefore, in such cases, it replaces the healthy tissues with the attributes of healthy tissues.\\
As a result, for every input image slice containing a candidate region(s), a new inpainted image will be synthesized. If the candidate region represents healthy anatomies, the inpainted model generates another healthy image, while for the tumoral regions, a tumor-free image will be synthesized. For each subject, a stack of volume is then reconstructed from the 2D inpainted slices. Finally, the residual volumes were calculated by finding the voxel-wise intensity differences between the original volume and the inpainted one. As the residual volume contains highlight regions presenting the tumor characteristics, it is named as the prior image. 
\begin{figure}[!ht]
\centering
\includegraphics[width=0.5\textwidth]{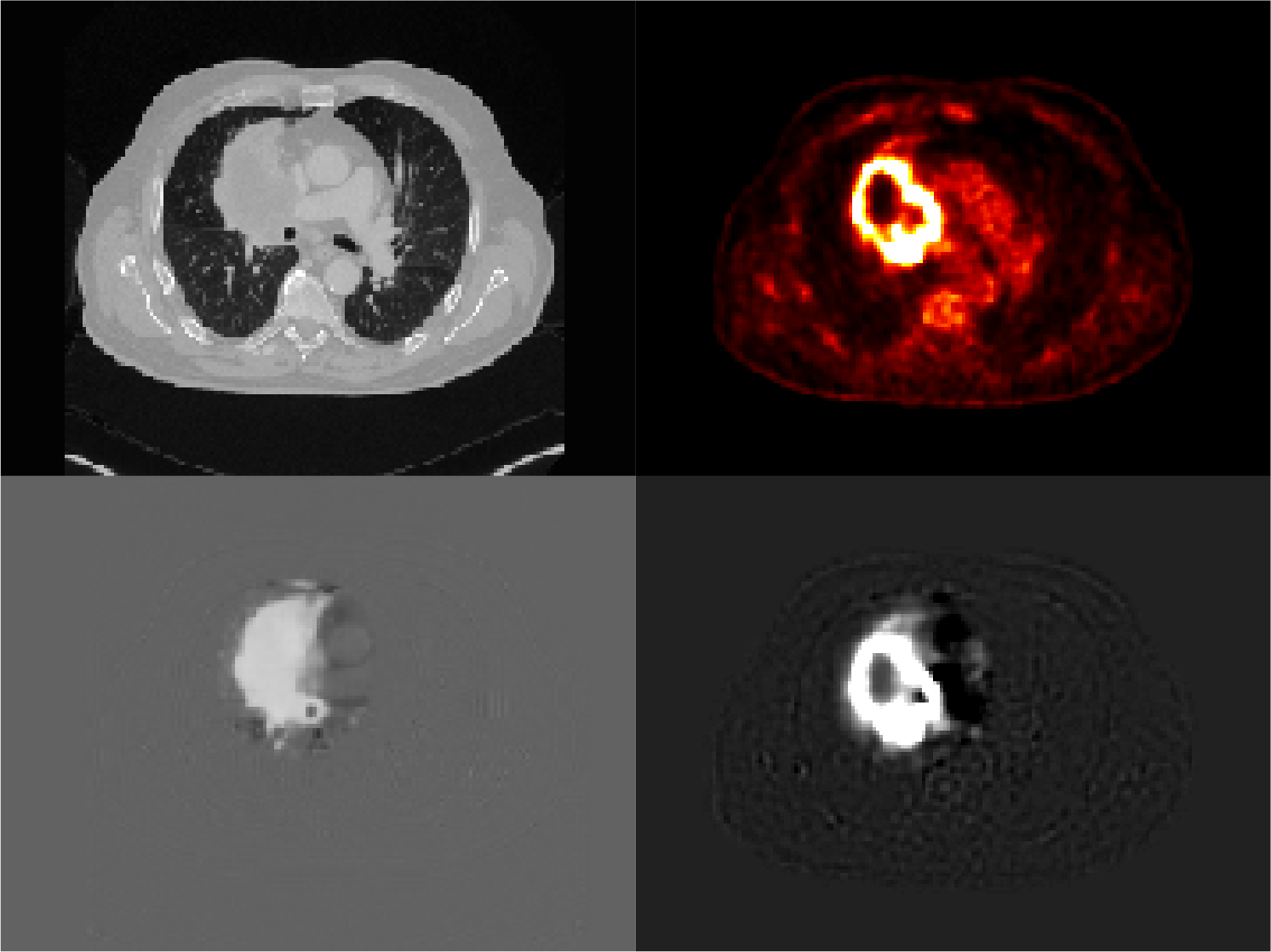}
\caption{Inpainting model residual outputs. Clockwise from top-left: 2D CT slice, 2D PET slice, 2D residual PET and 2D residual CT.}
\label{fig:residual_results}
\end{figure}
\subsection{Prior-aware segmentation model}
\begin{figure}[!ht]
\centering
\includegraphics[width=0.8\textwidth]{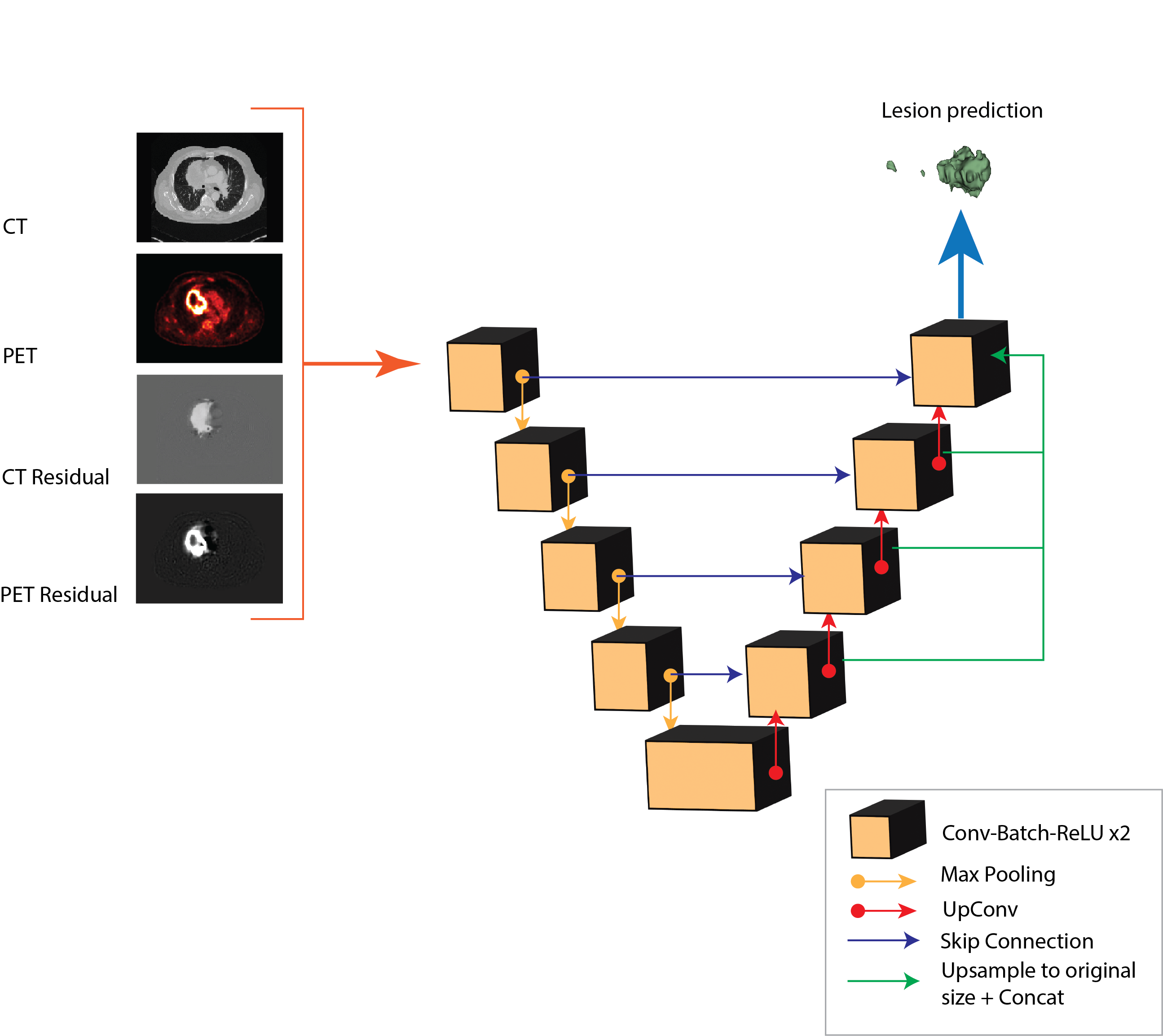}
\caption{U-Net architecture adopted for the lesion segmentation.}
\label{fig:unet}
\end{figure}

The second step of our proposed workflow combines the PET-CT volumes with the residual images obtained as describe in \ref{tumor_appearance}, in order to perform a segmentation task and predict the tumoral lesions (Figure \ref{fig:unet}). 

\subsection{Training procedures}
The inpainting model was trained with the same protocols described in \cite{astaraki2022prior}\cite{liu2018image}. In brief, this model was trained for 150 epochs in the first phase and another 150 epochs in the second phase. The differences between the first and second phases were the batch normalization flag which was deactivated in the encoder part of the model for the second phase, and the learning rate, which was decreased from 0.0001 to 0.00005 from the first to the second phase. \\
The images were corrupted by the random holes with variable sizes, which, on average, corrupted 25 to 30 percent of the image size. Since the healthy slices were used for training and the pathological slices were saved for the testing phase, no cross-validation was needed. Finally, in the test phase, a circular hole with a radius of 17 pixels was used to mask out the tumoral region for tumor inpainting.

For the prior-aware segmentation step, we adopted the conventional nnUNet framework \cite{Isensee2020}, with a 4-channel input (PET, CT, residual-PET and residual-CT), and a single channel output (tumor lesion mask). We trained the model for 1000 epochs, with 250 iterations per epoch and a batch size of 2. The chosen loss function is the default combined Dice-Cross Entropy Loss, with a learning rate of 0.0001 and the Adam optimizer.

\section{Data preprocessing and augmentation}
\subsection{Tumor appearance estimation}
In the first step of our proposed workflow, we aim at estimating the tumor lesion appearance on the 2D image slices. In order to generate the 2D image slices, the 3D PET and CT volumes are sliced along the axial orientation and normalized within the [0-1] range:  

\begin{math}
\newline
CT_{norm} = \frac{ CT_{x,y}-(-1000)}{800-(-1000)}\newline
\newline
 CT_{norm} = \begin{cases}
    0      & \quad \text{if } CT_{norm} <  0\\
    1      & \quad \text{if } CT_{norm} >  1\\
    CT_{norm}  & \quad \text{otherwise }
  \end{cases}
  \newline
  \newline
  PET_{norm} = \frac{ PET_{x,y}}{12}\newline
  \newline
PET_{norm} = \begin{cases}
    0      & \quad \text{if } PET_{norm} <  0\\
    1      & \quad \text{if } PET_{norm} >  1\\
    PET_{norm}  & \quad \text{otherwise }
  \end{cases}
  \newline
\end{math}

The normalized 2D slices are then stacked together to form an RGB image (Fig. \ref{fig:inpainting}). In detail, the normalized CT slice is used for the red and the blue channel, while the normalized PET slice is stacked as the green channel.
\subsection{Prior-aware segmentation}

In the second step of the proposed workflow, the voxel intensities for both CT and SUV PET volumes are clipped within the percentile range [0.5-99.5], followed by a z-score normalization:
\newline
\newline
\begin{math}
 CT_{x,y,z} = \begin{cases}
    CT_{0.5p}      & \quad \text{if } CT_{x,y,z} <  CT_{0.5p}\\
    CT_{99.5p}      & \quad \text{if } CT_{x,y,z} >  CT_{99.5p}\\
    CT_{x,y,z}  & \quad \text{otherwise }
  \end{cases}
  \newline
PET_{x,y,z} = \begin{cases}
    PET_{0.5p}      & \quad \text{if } PET_{x,y,z} <  PET_{0.5p}\\
    PET_{99.5p}      & \quad \text{if } PET_{x,y,z} >  PET_{99.5p}\\
    PET_{x,y,z}  & \quad \text{otherwise }
  \end{cases}
  \newline
  \newline
\newline
CT_{norm} = \frac{ CT_{x,y,z}-\mu_{CT}}{\sigma_{CT}}\newline
\newline
PET_{norm} = \frac{ PET_{x,y,z}-\mu_{PET}}{\sigma_{PET}}
\end{math}
\newline
\newline
As visually explained in Fig. \ref{fig:inpainting} both PET and CT residual volumes have intensity values in the range [-1,1]; for this reason, no normalization is performed for these volumes.

The default nnUNet data augmentation setting, presented in \cite{Isensee2020}, is adopted; the default configuration consist of random rotations, random scaling, random elastic deformations and gamma correction augmentation.

\section{Results}
In order to evaluate the performance of our proposed approach, we performed a 5-fold cross-validation on the 1014 PET-CT volumes of the AutoPET 2022 challenge.

The presented evaluation metrics are the same used in the challenge to determine the participant's rankings: Dice score, False Positive Volume and False Negative Volume (expressed in liters).

To better understand the lesion segmentation performances in relation to the tumor type (lung cancer, lymphoma, melanoma), in Table \ref{tab:dice} we provide the class-wise average scores, alongside the average negative (no lesions in the ground truth mask) and positive classes.

\begin{figure}[!ht]
\centering
\includegraphics[width=0.9\textwidth]{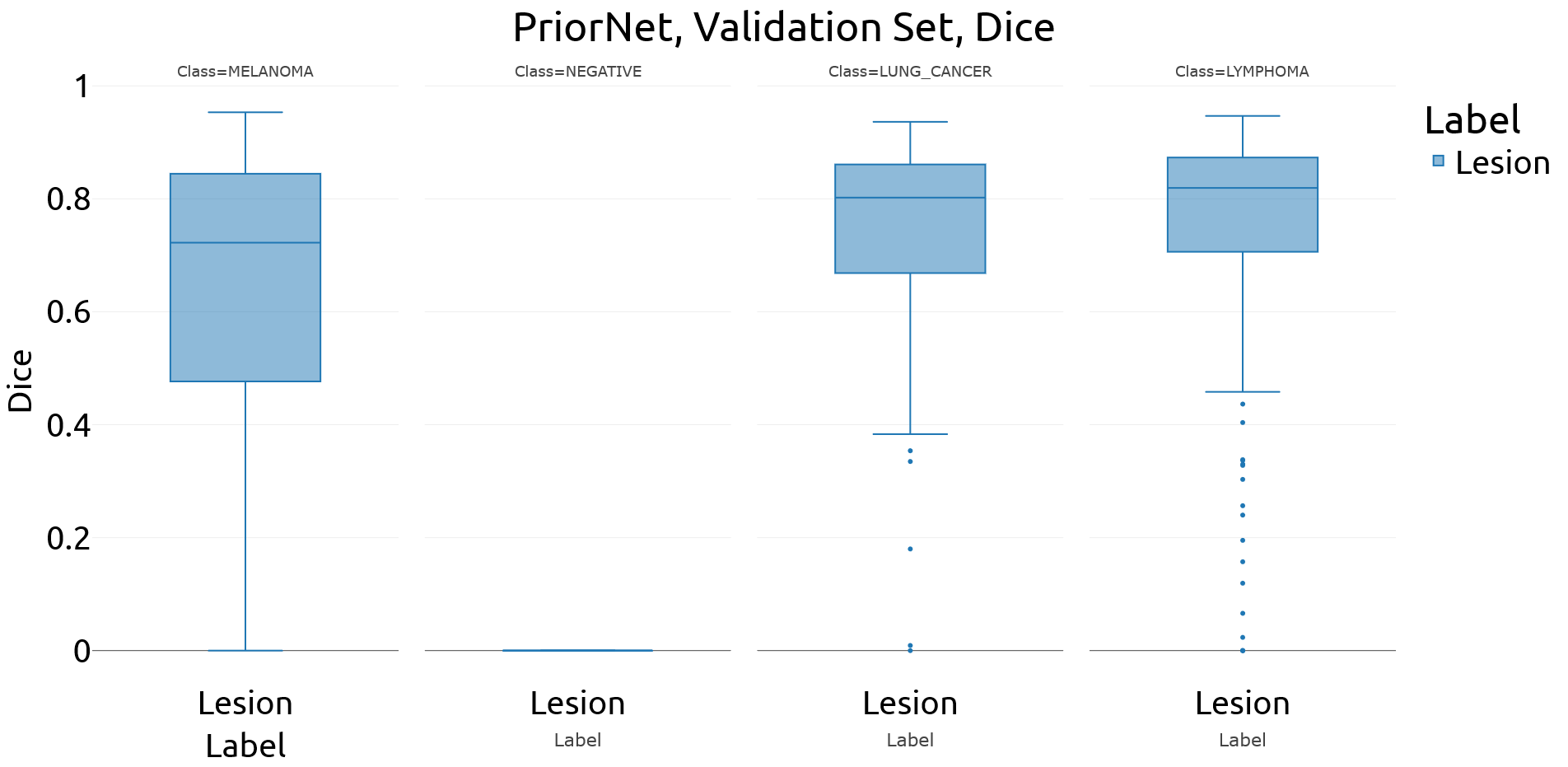}
\caption{PriorNet validation: class-wise Dice score}
\label{fig:results_dice}
\end{figure}

\begin{table}[!t]
\renewcommand{\arraystretch}{1.3}
    \caption{PriorNet validation: class-wise average metrics.}
    \label{tab:dice}
    \centering
    \begin{tabular}{|c||c|c|c|c|c|c|}
    \hline
    Metric & Global & Positive & Negative & Lung Cancer& Lymphoma & Melanoma\\
   
    \hline
    \hline
       Dice &-&	0.701&-&	0.743	&0.734& 0.638
\\
       \hline
False Positive Volume [ L ]&12.5&9.28&15.6&5.67&10.5&11.55
        
\\
         \hline
         False Negative Volume [ L ]&-&6.8&-&9.31&4.91&6.03
        
\\
\hline
    \end{tabular}

\end{table}
In Figure \ref{fig:results_dice} the Dice score distribution for the different tumor classes is represented.
\section{Public Code Repository}
The original code of PriorNet can be found at the following GitHub public repository: \url{https://github.com/SimoneBendazzoli93/PriorNet} 
\newpage
\bibliographystyle{splncs04}
\bibliography{bibliography}

\begin{thebibliography}{1}
\providecommand{\url}[1]{\texttt{#1}}
\providecommand{\urlprefix}{URL }
\providecommand{\doi}[1]{https://doi.org/#1}

\bibitem{astaraki2022prior}
Astaraki, M., Smedby, {\"O}., Wang, C.: Prior-aware autoencoders for lung
  pathology segmentation. Medical Image Analysis p. 102491 (2022)

\bibitem{brain_segmentation}
Bukhari, S.T., Mohy-ud Din, H.: E1d3 u-net for brain tumor segmentation:
  Submission to the rsna-asnr-miccai brats 2021 challenge (2021).
  \doi{10.48550/ARXIV.2110.02519}, \url{https://arxiv.org/abs/2110.02519}

\bibitem{Isensee2020}
Isensee, F., Jaeger, P.F., Kohl, S.A.A., Petersen, J., Maier-Hein, K.H.:
  {nnU}-net: a self-configuring method for deep learning-based biomedical image
  segmentation. Nature Methods  \textbf{18}(2),  203--211 (Dec 2020).
  \doi{10.1038/s41592-020-01008-z},
  \url{https://doi.org/10.1038/s41592-020-01008-z}

\bibitem{liu2018image}
Liu, G., Reda, F.A., Shih, K.J., Wang, T.C., Tao, A., Catanzaro, B.: Image
  inpainting for irregular holes using partial convolutions. In: Proceedings of
  the European conference on computer vision (ECCV). pp. 85--100 (2018)

\bibitem{ronneberger2015unet}
Ronneberger, O., Fischer, P., Brox, T.: U-net: Convolutional networks for
  biomedical image segmentation (2015)

\bibitem{ZHAO2020100357}
Zhao, W., Jiang, D., {Peña Queralta}, J., Westerlund, T.: Mss u-net: 3d
  segmentation of kidneys and tumors from ct images with a multi-scale
  supervised u-net. Informatics in Medicine Unlocked  \textbf{19},  100357
  (2020). \doi{https://doi.org/10.1016/j.imu.2020.100357},
  \url{https://www.sciencedirect.com/science/article/pii/S2352914820301969}

\end{thebibliography}
\end{document}